\documentclass{iau}
\usepackage{graphicx}

\title[SED Machine] 
{The SED Machine: A Spectrograph to Efficiently Classify Transient Events Discovered by PTF}

\author[Ngeow \etal]   
{Chow-Choong Ngeow$^1$, Nick Konidaris$^2$, Robert Quimby$^3$, Andreas Ritter$^1$, Alexander R. Rudy$^{1,5}$, Edward Lin$^1$ \and Sagi Ben-Ami$^4$}

\affiliation{
$^1$Graduate Institute of Astronomy, National Central University,\\ 
Jhongli City, 32001, Taiwan. 
\\ {\tt cngeow@astro.ncu.edu.tw} \\[\affilskip]
$^2$Cahill Center for Astronomy and Astrophysics, California Institute of Technology,\\ 
1200 E California Blvd., Pasadena, CA 91125, USA \\
$^3$Kavli-Institute for the Physcs and Mathematics of the Universe, University of Tokyo,\\
Kashiwanoha 5-1-5, Kashiwa-shi, Chiba, Japan \\
$^4$Department of Particle Physics and Astrophysics, The Weizmann Institute of Science,\\
Rehovot 76100, Israel \\
$^5$Fulbright Student Fellow, Taiwan
}

\pubyear{2012}
\volume{290}  
\pagerange{{\bf XXX--XXX}}
\jname{Feeding compact objects: Accretion on all scales}
\editors{T. Belloni, M. Mendez, C. Zhang \& S. Zhang, eds.}
\begin{document}

\maketitle

\begin{abstract}

The Palomar Transient Factory (PTF) is a project aimed to discover transients in the Universe, including Type Ia supernovae, core-collapse supernovae, and other exotic and rare transient events. PTF utilizes the Palomar 48-inch Telescope (P48) for discovering the transients, and follow-up mainly by the Palomar 60-inch Telescope (P60, for photometric light and color curves), as well as other telescopes. The discovery rate of PTF is about 7000 candidate transients per year, but currently only about 10\% of the candidates are being followed-up and classified. To overcome this shortcoming, a dedicated spectrograph, called the SED Machine, is being designed and built at the California Institute of Technology for the P60 Telescope, aiming to maximize the classification efficiency of transients discovered by PTF. The SED Machine is a low resolution ($R\sim100$) IFU spectrograph. It consists of a rainbow camera for spectrophotometric calibration, and a lenslet array plus 3-prism optics system for integrated field spectra. An overview of the science and design of the SED Machine is presented here. 

\keywords{instrumentation: spectrographs --- surveys --- supernovae: general}
\end{abstract}

\firstsection 
\section{Introduction}

\begin{figure}[b]
\begin{center}
 \includegraphics[width=4.0in]{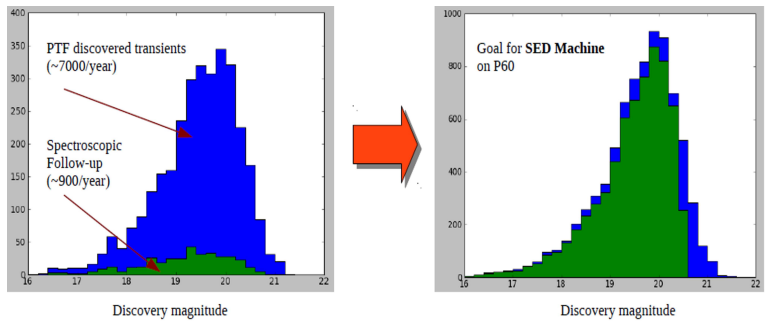} 
 \caption{Left panel shows the histogram of PTF discovered transients (including supernovae). The discovery rate for PTF is about 7000 per year, however only about 10\% of them being follow-up and classify. The goal of SED Machine is efficiently classify the discovered transients, bringing the classification efficiency to more than 90\% (right panel).}
   \label{fig1}
\end{center}
\end{figure}

The main goals for Palomar Transient Factory (\cite[PTF, Law \etal\ ~2009; Rau \etal\ ~2009]{law,rau}) include finding Type Ia and core collapse supernovae, discovering various rare transient events, as well as dedicated time domain study (such as the PTF-Orion project). The PTF employs a two-telescope approach at the Palomar Observatory: the P48 Telescope serves as discovery and survey telescope, while the P60 Telescope is used for multi-colors light curves follow-up of the discovered, or interesting, explosive transients. The P48 Telescope is equipped with a wide field mosaic CCD imager for $\sim7$ deg$^2$ field-of-view (FOV), and observes in single band filter (either $g$, $R$ or $H\alpha$). Image subtraction techniques are used to find transients (in real time) from the images taken from P48. About 80\% and 50\% of P48 and P60 time is allocated to PTF, respectively. Other medium to large aperture telescopes within PTF Network are also used for follow-up observations (mainly spectroscopic), including P200 Telescope, Keck Telescopes and others, for classification of transients discovered by PTF. Even with a relatively large amount of telescopes time, only $\sim10$\% of the transients can be followed up (see Figure \ref{fig1}). To catch up with the PTF discovery rate, a new approach must be adopted  --- the SED Machine is designed to observe most of the transients.

\section{The SED Machine}

SED Machine (for further details, see \cite[Ben-Ami \etal\ ~2012]{benami}) is a low resolution integral field unit (IFU) spectrograph coupled with a Rainbow Camera (RC), and it is planned to be installed at P60 in 2013. The IFU spectrograph utilizes a lenslet array and a prism to produce $\sim3600$ spectra packed at CCD plane, each with resolution of $R\sim100$. The RC is a large format CCD with different filters at 4 quadrants, in order to provide spectrophotometric calibration to within $\sim5$\% accuracy.

How can the SED Machine improve the classification efficiency? Study (\cite[Ben-Ami \etal\ ~2012]{benami}) has shown that spectral resolution of $R\sim100$ is adequate to classify most types of transients, hence allowing shorter exposure times at small aperture telescopes. The use of triple prisms permits a large wavelength coverage (370 to 920 nm) and high throughput, and yet providing $\sim$constant spectral resolution. The acquisition and pointing efficiency can be improved by adopting lenslet-based IFU. The RC provides calibration efficiency for monitoring atmospheric extinction using a relatively large number of stars within its FOV. Finally, CCD detectors with high efficiency and low noise permit the observation and classification of $\sim18$ transients per night.

\section{Future Implications}

The SED Machine is designed to be adopted for other 2 to 4 meter class telescopes, and not just on P60. In addition to transients classification, SED Machine may also be used for asteroid taxonomy study, galaxy clusers redshift determination, stellar principal parameters (surface gravity, $T_{eff}$ \& $[M/H]$) estimation, and other science.

\end{document}